\documentclass[amsmath,nofootinbib]{revtex4}

\usepackage[dvips]{graphicx}
\usepackage{dcolumn}
\usepackage{bm,xcolor}
\usepackage{footmisc}

\newcommand*{\no}{\noindent}
\newcommand*{\bea}{\begin{eqnarray}}
\newcommand*{\eea}{\end{eqnarray}}
\newcommand*{\be}{\begin{equation}}
\newcommand*{\ee}{\end{equation}}

\newcommand*{\pref}[1]{(\ref{#1})}

\newcommand*{\prefr}[2]{(\ref{#1}-\ref{#2})} 
\newcommand*{\nn}{\nonumber}

\newcommand*{\tr}{\mathrm{tr}}

\graphicspath{{./}{publPlots/}}

\begin{document}


\title{Field theory as a tool to constrain new physics models}

\author{Axel Maas}
\email{axel.maas@uni-graz.at}
\affiliation{Institute of Physics, University of Graz, Universit\"atsplatz 5, A-8010 Graz, Austria}

\date{\today}

\begin{abstract}

One of the major problems in developing new physics scenarios is that very often the parameters can be adjusted such that in perturbation theory almost all experimental low-energy results can be accommodated. It is therefore desirable to have additional constraints. Field-theoretical considerations can provide such additional constraints on the low-lying spectrum and multiplicities of models. Especially for theories with elementary or composite Higgs particle the Fr\"ohlich-Morchio-Strocchi mechanism provides a route to create additional conditions, though showing it to be at work requires genuine non-perturbative calculations. The qualitative features of this procedure are discussed for generic 2-Higgs-doublet models, grand-unified theories, and technicolor-type theories.

\end{abstract}

\maketitle

\section{The situation in the standard model}

One of the most basic requirements of particle physics theories is that their experimentally observable consequences are gauge-invariant. In QCD confinement dictates that gauge-invariant bound states are the only observable states\footnote{Often the top is referred to as an observable quark. However, this is not correct, as also here the color charge has to be dressed in some way to obtain a gauge-invariant state. This is only possible with some light quark, but has effects only of the order of $\Lambda_\text{QCD}$.}. The situation in the electroweak sector of the standard model is more subtle. In electrodynamics, as an Abelian sector, gauge-invariance can be achieved by a suitable dressing \cite{Haag:1992hx}. This yields gauge-invariant states and a gauge-invariant electric charge. For the non-Abelian weak interactions, this is no longer the case \cite{Haag:1992hx}. In principle, the weak interactions would behave like QCD. However, the Brout-Englert-Higgs (BEH) mechanism leads to a different realization.

The BEH effect hides the gauge symmetry. However, this by no means in itself directly influences the physical spectrum. In fact, its textbook realization in terms of a vacuum expectation value \cite{Bohm:2001yx} is gauge-dependent \cite{Lee:1974zg}. A gauge-invariant, and necessarily non-perturbative, formulation does not require any vacuum expectation value \cite{Frohlich:1980gj,Maas:2013aia}. This is manifest in the lattice-regularized version of the Higgs sector, where there is no qualitative distinction between a QCD-like region exhibiting QCD-like confinement and a region exhibiting the BEH effect \cite{Osterwalder:1977pc,Fradkin:1978dv,Caudy:2007sf,Maas:2013aia}.

This leaves open the question of what the observable states of the theory are. The connection to the QCD-like region implies these should be bound states. Furthermore, only bound states can be gauge-invariant, and thus observable particles\footnote{The Nielsen identities are sometimes invoked to show the gauge-invariance of some masses, but they only guarantee gauge-parameter-invariance, which is a weaker statement. A counter example is already given in the original work \cite{Nielsen:1975fs}. } \cite{Frohlich:1980gj,Banks:1979fi,'tHooft:1979bj}. Nonetheless, the description in terms of the gauge-dependent elementary degrees of freedom in perturbation theory yields very convincing results \cite{Bohm:2001yx,pdg}, as if these would be observable.

This apparent contradiction is resolved by the Fr\"ohlich-Morchio-Strocchi (FMS) mechanism \cite{Frohlich:1980gj,Frohlich:1981yi}. In a nutshell, this proceeds as follows: Gauge-invariant states can be formulated as composite operators. Choose a gauge with non-vanishing Higgs expectation value. Take then the composite operators in the quantum number channels desired and expand the Higgs field $\phi$ around its expectation value $vn^i$, $\phi^i(x)=\eta^i(x)+n^iv$, with $n^i$ some constant isospin vector. For the $0^+$ custodial-singlet channel this yields \cite{Frohlich:1980gj,Frohlich:1981yi}
\bea
&&\langle\phi_i^\dagger(x)\phi^i(x)\phi_j^\dagger(y)\phi^j(y)\rangle\nn\\
&&\approx v^4+4v^2(\zeta+\langle\eta^\dagger_i(x) n^i n_j^\dagger\eta_j(y)\rangle)+{\cal O}(\eta^3/v^3)\label{correl},
\eea
\no and for the $1^-$ custodial-triplet channel
\bea
&&\langle(\tr\tau^a\varphi^\dagger D^\mu\varphi)(x)(\tr\tau^a\varphi^\dagger D_\mu\varphi)(y)\rangle\nn\\
&&\approx \tilde{\zeta}\langle W_i^\mu W^i_\mu\rangle+{\cal O}(\eta W/v)\label{correl2}.
\eea
\no Herein $\zeta$ and $\tilde{\zeta}$ are some constants, $\varphi$ an SU(2)-valued representation of the direction of $\phi$ in isospin space \cite{Shifman:2012zz}, and the $\tau^a$ are generators of the custodial symmetry  \cite{Shifman:2012zz}. The latter makes the vector operator a custodial triplet, rather than a gauge triplet. Thus, the masses determined by the poles of the correlators on both sides have to coincide to this order, yielding the equivalence of the mass spectrum in terms of gauge-dependent and gauge-invariant degrees of freedom. It is also remarkable that in \pref{correl2} the custodial triplet on the left-hand-side is turned into a gauge triplet on the right-hand-side. This explains that both the gauge-dependent and the gauge-invariant spectrum coincides, and why the standard description of the Higgs sector of the standard model is so successful. It is tempting to identify the bound states with the Higgs and $W$/$Z$ also by name. This will not be done here, as a careful distinction will be essential.

However, this requires two conditions to be fulfilled. One is that the expansion in \prefr{correl}{correl2} is valid. This requires an explicit calculation of both sides of the equations, a task which requires non-perturbative methods, e.\ g., using lattice calculations\footnote{These neglect QED, QCD, and fermions. This is not expected to make a qualitative impact \cite{Frohlich:1980gj,Frohlich:1981yi}, but can make a quantitative one.}. All results to date indicate that the range of validity is restricted to Higgs masses above the $W$ mass \cite{Evertz:1985fc,Langguth:1985dr,Wurtz:2013ova,Maas:2013aia,Maas:2014pba}, in contrast to the perturbative range of validity of the BEH effect \cite{Bohm:2001yx}. Towards larger Higgs masses, the limit is not yet fully established \cite{Wurtz:2013ova,Maas:2014pba}, but indications exist that the upper limit of validity may be lower than expected from perturbative considerations of the BEH effect \cite{Maas:2014pba}. The second requirement is that in \prefr{correl}{correl2} the multiplicities are such that they agree with experiment. In the standard model, this is guaranteed by the coincidence of the custodial and the gauge group. Thus, in the standard model this appears to be the mechanism establishing a gauge-invariant spectrum and justifying the use of perturbation theory to calculate observables, up to sub-leading corrections in the FMS expansion.

Besides those necessary requirements, there are also others which must be met to explain the success of perturbation theory. One is that in other channels no bound states appear which would be inconsistent with experiment. In the FMS mechanism this is guaranteed, as to leading order other channels just create the scattering states known in perturbation theory. Also in this respect, lattice results support the FMS mechanism \cite{Wurtz:2013ova,Maas:2014pba}.

\section{Beyond the standard model}

The FMS mechanism explains why the usual perturbative description of the observable spectrum in the standard model works. The non-perturbative calculations justify that this is the case, and that the expansion in \prefr{correl}{correl2} is a valid one. Of course, at some point deviations should appear, but they may easily be beyond the experimental reach for some time to come. E.\ g., there should be a finite size of the observed states. Calculating and measuring them would be a major challenge, but a coincidence would ultimately establish our understanding of the Higgs sector from a field-theoretical perspective, vindicating the combination of the BEH effect and the FMS mechanism as the language to describe the electroweak sector. Higher-order calculations of, e.\ g.\ \prefr{correl}{correl2} and comparison to perturbation theory should provide insight where deviations can be expected.

However, in beyond-the-standard-model (BSM) scenarios, the situation is far less clear. The purpose of this paper is to point out some possible conflicts.

\subsection{2-Higgs-doublet models}

2-Higgs-doublet models (2HDMs) \cite{Lee:1974jb,Branco:2011iw} enjoy an amount of popularity both as stand-alone models, but especially as low-energy effective models of, e.\ g., supersymmetric extensions of the standard model. There is a large multitude of these models, but ultimately they are characterized by how their enlarged custodial group is broken both explicitly and spontaneously. In case of only one of the doublets showing a condensation, the situation is very similar to the standard model. Of course, to establish that again the FMS mechanism works would require a non-perturbative calculation. Given the possible narrow range of masses for which it is applicable, this would be an important test. Lattice calculations of this question are possible \cite{Lewis:2010ps,Maas:2014nya}, but tend to be quite expensive due to the large parameter space.

However, if both doublets condense, but maintain individual custodial symmetries within each doublet, an additional effect arises \cite{Maas:2014nya}. For the (pseudo)scalar channels an expansion of type \pref{correl} yields an equivalence between the various Higgs bound states with their custodial quantum numbers and the corresponding elementary Higgs. Only the expansion for the scalar uncharged channel requires some comments. Using the two Higgs fields $h$ and $H$ with condensates $v_h$ and $v_H$ yields
\bea
&&\langle h_i^\dagger(x) h^i(x)h_j^\dagger(y)h^j(y)\rangle\nn\\
&&\approx v_h^4+4v_h^2(c+\langle\eta_{hi}^\dagger(x) n_h^i n_{hj}^\dagger\eta_h^j(y)\rangle)+{\cal O}(\eta_h^3)\nn\\
&&\langle H_i^\dagger(x) H^i(x)H_j^\dagger(y)H^j(y)\rangle,\nn\\
&&\approx v_H^4+4v_H^2(c+\langle\eta_{Hi}^\dagger(x) n_H^i n_{Hj}^\dagger\eta_H^j(y)\rangle)+{\cal O}(\eta_H^3)\nn.
\eea
\no Since the two states on the left-hand-side have the same quantum numbers, the different expansions on the right-hand-side imply the existence of (at least) two poles, and this recovers the two scalar particles of the perturbative language. Of course, a cross-expectation value between $h$ and $H$ on the left-hand side is also possible, but this will yield a mixed two-point function between $h$ and $H$ on the right-hand side, which vanishes, and only higher orders contribute. Thus, there are two distinct scalar isoscalar Higgs particles in the spectrum to this order.

More interesting is the case of the vector particles living inside two distinct custodial triplets,
\bea
&&\langle(\tr\tau^a\varphi_h^\dagger D^\mu\varphi_h)(x)(\tr\tau^a\varphi_h^\dagger D_\mu\varphi_h)(y)\rangle\nn\\
&&\approx \tilde{d}\langle W_i^\mu W^i_\mu\rangle+{\cal O}(\eta_h W/v_h),\nn\\
&&\langle(\tr\omega^a\varphi_H^\dagger D^\mu\varphi_H)(x)(\tr\omega^a\varphi_H^\dagger D_\mu\varphi_H)(y)\rangle\nn\\
&&\approx \tilde{f}\langle W_i^\mu W^i_\mu\rangle+{\cal O}(\eta_H W/v_H)\nn
\eea
\no where $\tau$ and $\omega$ are now generators of the two distinct custodial symmetry groups, and the $\tilde{d}$ and $\tilde{f}$ are some constants. In both cases the same correlation function appears on the right-hand-side, the one of the $W$. Thus there are six, instead of three, particles with the same mass as the $W$ bosons. Since this multiplicity is in contradiction with experiment \cite{pdg}, this would immediately rule out these models. However, this argument requires the FMS mechanism to be at work for both Higgs doublets, and this in turn requires a verification of all these relations non-perturbatively.

This is a first example of how the FMS mechanism can be used to constrain, or possibly even rule out, models. Of course, this requires to explicitly check whether the FMS mechanism for a given set of parameters works. But if it does not work than there is little reason to believe that the gauge-invariant states and the elementary particles should have the same mass spectrum. In this case, a perturbative description would likely fail to give an adequate description of the experimentally observed masses. Insisting that only gauge-invariant states should appear in the spectrum then requires a non-perturbative calculation anyhow.

\subsection{Grand-unified theories}

A particular interesting case for the FMS mechanism are grand-unified theories (GUTs) \cite{Bohm:2001yx}, often invoked to establish a connection between various essentially unrelated but linked features in the standard model, like the electric charges. In such a theory, the whole standard-model gauge group emerges from a single gauge group at the GUT scale. There, it is broken by some BEH effect to the standard model group, yielding heavy gauge bosons $X$, leaving the standard-model gauge bosons $W$ massless. A second BEH effect with a second set of Higgs fields then provides the standard model breaking. Of course, both Higgs fields are charged under the full gauge group.

From the FMS point of view the theory is a modified version of the 2HDM, though potentially with both Higgs doublets in different representations. Both doublets again require a BEH effect to occur, though at vastly different scales. For the scalar particles the same arguments can be made as for the 2HDMs, justifying the expectation of two Higgs particles in the spectrum.

The situation is quite different for the $1^-$ channel. Here the expansion for either Higgs field takes the form
\bea
&&\langle(\tr\tau^a\varphi^\dagger D^\mu\varphi)(x)(\tr\tau^a\varphi^\dagger D_\mu\varphi)(y)\rangle\nn\\
&&\approx t^{ij}(\langle W^i_\mu W_j^\mu\rangle+\langle X^i_\mu X_j^\mu\rangle)+{\cal O}(\eta W/v,\eta X/v )\label{wgut},
\eea
\no where the $\varphi$ is either of the Higgs fields, the $\tau$ are matrices in the respective custodial symmetry groups, the $W$ belong to the low-energy sub-group, and the $X$ are the GUT-scale additional gauge bosons. The tensor $t^{ij}$ is uniquely determined from the custodial and gauge group structure, as well as the breaking pattern. The $t_{ij}$ can be zero for some of the indices, and thus the right-hand side may contain all or only some of the gauge bosons. In this case, the right-hand side has two poles, one for each set of gauge bosons. Since another expression exists for the second Higgs fields, this requires a rather subtle group structure, such that $t^{ij}$ is of such a type as that the correct number of light standard-model gauge bosons of all types emerge. Considering only the weak interaction, exactly three states must remain. Hence, there must be a custodial triplet of light vector bosons. To have the same spectrum for the heavy bosons as in perturbation theory would further require a corresponding heavy multiplet. But since no experimental constraints exist for the heavy particles, this could be allowed to vary.

The upshot is, that the global and local group structure have to fit such that in \pref{wgut} the right number of light vector states remains. This number will in general differ from the number of light gauge bosons, and thus constrains the possible theories. This requires again the FMS to work, and thus its non-perturbative confirmation.

\subsection{Technicolor}

The situation for technicolor \cite{Andersen:2011yj}, which aims at replacing the Higgs by a bound-state of new fermions bound by technicolor as a new gauge interaction, is much more subtle, if one regards the theory's substructure\footnote{Ignoring extended sectors, as they will not alter the following qualitatively.}, rather than just its effective low-energy model. The latter is essentially similar to the standard model or the 2HDMs.

The full technicolor theory is, however, more challenging. Leaving once more aside QCD, QED, and the standard model fermions, the theory is essentially a set of fermions with two gauge interactions, one being the usual weak one. To be compatible with experiment, this theory's gauge-invariant spectrum must have as lightest particles a triplet of vectors and a scalar as the next excitation. This is a remarkable challenge. All theories with fermions and only one gauge interaction have so far found to be only of two types \cite{Gattringer:2010zz,Andersen:2011yj,Kuti:2014epa}: One is QCD-like with pseudoscalars as lightest particles. The other has a scalar as lightest particle. No theory with fermions and a vector as lightest particle is yet known. But precisely this is required for a suitable, gauge-invariant phenomenology. It is, however clear that a triplet can only be obtained if there exists a global (flavor) symmetry which permits triplet representations, at least as long as no further interactions should be introduced.

The FMS mechanism, in the sense of a mean-field approximation, cannot help in this instance, as can be seen directly for the scalar channel. Expanding the scalar operator $\langle(\bar{\psi}\psi)(p)(\bar{\psi}\psi)(-p)\rangle$ around the chiral condensate $\langle\bar{\psi}\psi\rangle$ does not yield the elementary fermion propagator, but just a scattering state. Likewise, no possibility exists for the vector channel. It is thus particular important to emphasize that trying to replicate the standard-model mechanism in a technicolor theory immediately faces the problem that the Higgs is not a gauge-invariant object as noted above. Hence, to have the technicolor theory provide the BEH effect, it is not sufficient to have, without the weak interaction, a light scalar. The relevant condition sine-qua-non is the presence of a gauge-invariant vector triplet as the lightest particle in the spectrum in the theory with both gauge fields, as well as a gauge-invariant scalar with a slightly higher mass.

Using, e.\ g., minimal walking technicolor \cite{Andersen:2011yj}, i.\ e.\ a technicolor sector of two colors with two flavors of adjoint quarks, the relevant operator would be a $\rho$-type operator $t^{a}_{ij}\bar\psi_i^r D_\mu^{rs}\psi_j^s$. It does not matter, which gauge field is inside the covariant derivative: Since both operators have the same quantum numbers, both should exhibit a pole at about 80 GeV, provided both have overlap with the ground state. In general, in fact, it should be expected that both operators mix. Similar, the next state must be a scalar, which can either be described by a gauge-boson-ball or by the usual scalar meson operators. Any further states in these channels, or in other channels, must then be far up in the spectrum or very weakly coupled to be compatible with present experimental bounds \cite{pdg}.

The calculation of the spectrum is a genuine non-perturbative question. It requires a gauge theory with fermions to have as lightest excitation a vector, something unseen so far. The requirement of gauge-invariance is far stronger than having a suitable low-energy model.

\section{Summary}

The demand of gauge invariance of physical observables must be taken directly as a demand on the spectrum of any theory. In case of the standard model, the FMS mechanism justifies that the spectrum can nonetheless be rather well described by the spectrum of the gauge-dependent elementary states, irrespective of using perturbation theory or non-perturbatively.

In proposals for BSM physics, a similar reasoning can be applied when a BEH effect is present. The FMS mechanism then gives conditions, especially on the group structure, whether a description in terms of gauge-dependent variables is possible at all. The applicability of the FMS mechanism is, however, a dynamical question, requiring genuine non-perturbative methods to answer. This potentially restricts the classes of theories compatible with the low-energy physics substantially.

In theories without elementary Higgs fields, and thus absence of a BEH effect, the situation is more complicated. Without the FMS mechanism, there is no simple way to determine the relation between the elementary states and the gauge-invariant states. The only exception may be in case of very heavy additional particles, where, just like with the top, the heavy degrees can be appropriately described without taking the dressing into account. But this will only give information on the heavy states of a theory. For the light states, there is no alternative to calculating the gauge-invariant mass spectrum. Taking technicolor seriously, this implies to search for gauge theories with fermions with lightest particles being vectors, something which has not been seen so far.

Closing, using field-theoretical consequences of gauge-invariance imposes additional, sometimes powerful, constraints on new physics scenarios beyond limitations on parameters. It restricts the particle content and has implications for the global symmetries. This will be a powerful tool to select among possible new physics scenarios in a time without strong experimental hints.\\

\no{\bf Acknowledgments}

I am grateful to R.\ Alkofer and H.\ Gies for a critical reading of the manuscript.

\bibliographystyle{bibstyle}
\bibliography{bib}


\end{document}